\documentclass[aps,prb,twocolumn]{revtex4}
\usepackage{tabularx}
\usepackage{bm}
\usepackage{euscript}
\usepackage{epsfig,psfrag,subfigure}
\usepackage{graphicx}
\usepackage{color}
\usepackage{amsfonts}
\usepackage{exscale}
\usepackage{amsbsy}
\usepackage{stmaryrd}
\usepackage{wasysym}

\def\be{\begin{equation}}       \def\ee{\end{equation}}
\def\bea{\begin{eqnarray}}      \def\eea{\end{eqnarray}}
\def\ba{\begin{array} }
\def\ea{\end{array} }
\def\bnum{\begin{enumerate} }
\def\enum{\end{enumerate}}

\def\=>{\Rightarrow}
\def\>{\rightarrow}

\def\eye2{Fathbb{I}}

\begin{document}
\title{One hole in the two-leg t-J ladder and adiabatic continuity to the non-interacting limit}

\author{S. R. White, D. J. Scalapino, and S. A. Kivelson}
\affiliation{Department of Physics, University of California, Irvine, California 92697, USA,\\
Department of Physics, University of California, Santa Barbara, California, 93106, USA,\\
Department of Physics, Stanford University, Stanford, California 94305, USA}
\begin{abstract}
We have carried out density-matrix-renormalization group (DMRG) calculations for
the problem of one doped hole 
in a two-leg $t-J$ ladder. Recent studies have
concluded that exotic ``Mott'' physics --- arising from the projection onto the space
of no double-occupied sites --- is manifest in this model system, leading to charge
localization and a new mechanism for charge modulation. In contrast, we show that
there is no localization and that the charge density modulation arises when
the minimum in the quasiparticle dispersion moves away from $\pi$.
Although singular changes in the quasiparticle dispersion do occur as a function of model parameters, all
the DMRG results can be qualitatively understood from a non-interacting
``band-structure'' perspective.
\end{abstract}
\date{\today }
\maketitle

A strongly correlated quantum system is one in which the interactions are at
least comparable to the kinetic energy so  weak-coupling, perturbative
approaches cannot be justified.  
However, a key question is -- under what circumstances does the behavior of
such systems extrapolate smoothly to the weakly interacting limit so that,  at
least at the phenomenological level, weak coupling intuitions can still be
applied?  There are certainly forms of broken symmetry, such as charge-density
wave order in more than 1D, which are at the very least unnatural at weak
coupling, and  there can be still more exotic phases, especially those that
support topological order and fractionalization, which  have no weak-coupling
analogues. What about the important case of a doped Mott insulator? 
It has been argued by many authors that there is an additional quantity,
sometimes referred to as ``Mottness'', which through the effect of the
constraint of no double-occupancy produced by a strong local ``Hubbard $U$,'' 
can invalidate the quasiparticle picture and preclude the adiabatic
continuation to 
the weakly interacting reference state that underlies Fermi
liquid theory. 

The idea that the quasiparticle picture fails qualitatively 
has gained strong support from a set of papers by Zhu {\it et al}\cite{Zhu1,Zhu2,Zhu3, Zhu4}, 
in which extensive numerical experiments have been carried out using the density matrix 
renormalization group (DMRG)\cite{dmrg} on a set of $t-J$ 
ladders.
It has long been thought  that the undoped two-leg $t-J$ ladder is adiabatically related to a band insulator, and
 a number of early exact diagonalization\cite{Troyer} and quantum Monte Carlo\cite{Brunner}  studies 
 supported the idea that doped holes 
 form conventional quasiparticles.  In striking contrast, Zhu  {\it et al} 
reported that a doped hole in a two leg ladder {\it localizes} at large length
scales, {\it a finding that is incompatible with Bloch's theorem for any
quasiparticle state.} Similar localization was reported on three and four leg systems, although 
the data is less extensive.  Zhu  {\it et al} 
proposed an explanation for this behavior based on considerations of ``hole phase-strings'' 
and a new type of ``Weng statistics.''
It has been further proposed,\cite{Zaanen} that this new paradigm 
can account for a wide range of phenomena in doped Mott insulators, 
including stripe formation in the cuprates.

In this paper, we have focussed on the  two-leg $t-J$ ladder with one doped
hole.  We have carried out DMRG calculations to extract the ground-state
properties of  ladders of length up to $L=1000$, and time-dependent DMRG\cite{tdmrg1,tdmrg2,tdmrg3}
(tDMRG) calculations on ladders up to $L=120$ to obtain unprecedentedly
complete information concerning the dynamical one-hole Green function, $G$.
Following Zhu {\it et al} we have considered  a range of values of the
parameter $\alpha$, the ratio of the hopping matrix elements and the exchange couplings on the legs and
the rungs of the ladder.  In contrast to them, we find that the one hole state
is never localized. On the other hand, we corroborate their discovery that a
notable change in the character of the one-hole state  occurs at a critical
value of $\alpha=\alpha_c\approx 0.68$; in particular  the quasiparticle
effective mass diverges as $\alpha\to \alpha_c$.  However, this singular
behavior does not imply the existence of a phase transition, as changes in the
properties of a single doped hole do not reflect changes in the thermodynamic
state of the system.  Indeed, we show directly from the structure of $G$ that
the quasiparticle is well defined for  $\alpha$ on both sides of $\alpha_c$, that 
there is no   ``spin-charge separation,''
and that the quasiparticle weight, $Z(\alpha)$, is  always substantial.  Indeed,
all the properties of the low energy one hole states can be adiabatically
related to those of a single hole in a non-interacting ``band'' insulator --
the singular changes reflect a shift of the ground-state sector  from a Bloch
wave vector $k=\pi$ for $\alpha < \alpha_c$ to $k =k_0(\alpha) <\pi$ for
$\alpha>\alpha_c$.  The divergent effective mass 
dramatically reflects a point at
which the minimum of the quasihole dispersion, $\varepsilon(k)$, shifts 
away from
$\pi$. 

In this paper we will study the 2-leg $t-J-\alpha$ model 
\begin{equation}
  H= -\sum_{\langle i,j\rangle ,\sigma}t_{ij}c^{\dagger}_{i,\sigma}c_{j\sigma}+\sum_{\langle i,j\rangle }(\bold S_i\cdot \bold S_j-\frac{1}{4}n_i n_j) .
\label{eq:1}
\end{equation}
Here $\langle ij\rangle $
indicates nearest-neighbor sites with $t_{ij}=t $ and $J_{ij}=J$ on the rungs, and
 $t_{ij}=\alpha t $ and $J_{ij}=\alpha J$ on the legs, 
$c_{j,\sigma}^\dagger$ creates an electron on site $j$ with spin polarization
$\sigma$,  the spin operator on site $j$ is $\bold S_j$, the charge is
$n_j=\sum_{\sigma}c_{j,\sigma}^\dagger c_{j,\sigma}$, and the action of the
Hamiltonian is restricted to the Hilbert space with no doubly occupied sites,
$n_j=0$, 1.  The index $ i=
(l_x,l_y)$ with $l_y=1 $ and 2 denoting the two legs and $ l_x$
 runs from 1 to L. This is the same realization of the $t-J$ model that was studied
by Zhu {\it et al.} for a range of $\alpha$ with $J/t=1/3$.  They gave a 
quasiparticle interpretation to their results for $\alpha < \alpha_c\approx
0.7$, but they identified a transition  at $\alpha=\alpha_c$, such that, among
other anomalies,  for $\alpha > \alpha_c$ and ladders of length $L > 100$, they
reported 
 localization of the charge in a region  of width $\xi \sim 100 < L$.

Our ground state DMRG calculations were fairly standard, the main exception being 
that an unusually large number of sweeps were needed for the one hole ground states. 
All the calculations reported here were
performed using the ITensor library (http://itensor.org). 
A sufficient number of states, roughly 200-400 for the one hole case, were kept
to limit the truncation error per step to $\sim 10^{-10}$.  For each system,
first the ground state for the undoped system was obtained, with four sweeps
giving high-accuracy convergence, and this matrix product state $|\phi \rangle$
was stored. We then applied the operator $c_{j_0\downarrow}$, where $j_0$
is a site at the center of the system, creating a one hole state with the hole localized 
in the center. Sweeps were then carried out, resulting in a set of ever better
approximate one-hole groundstates, $|\psi(s)\rangle$, where $s$ 
indicates the number of sweeps.  At each sweep we made diagonal measurements 
of the energy and the density on each site,
as well as off-diagonal measurements of 
the hole amplitude, $F(j,s) = \langle \phi | c^\dagger_{j\downarrow} |\psi(s)\rangle$. 

Figure 1(a) shows the spreading of the density in a $1000\times2$ system versus sweep 
with $\alpha = 1$.  Here the hole density for site $j$ is $n_h(j) \equiv 1 - n_j$; the figure shows
the rung hole density $\bar n_h(l_x) = \sum_{l_y} n_h(l_x,l_y)$.
The density continues to spread out as the sweeps progress. (Note, to facilitate  comparisons with  previous results, 
we have eschewed  tricks that could be used to accelerate convergence to the true ground state, such as starting with 
a delocalized hole as the initial state.) The inset in Fig.~\ref{fig:1}(a) 
shows
the full width at half-maximum (FWHM) of the charge density profile for
$\alpha=1$ ladders of different lengths $L$.  This value of $\alpha$ is greater
than $\alpha_c$ and places the system in the region where Zhu {\it et al.} reported
localization. However, as seen in the inset, we find that the FWHM scales as L.  The saturation
of the FMHW reported by Zhu {\it et al.}  in Fig. 2c of Ref [4] appears to be an artifact of their calculation 
which arises from limiting the number of  DMRG sweeps. In fact, as shown in Fig. 4c of  [4],
they, too, find
 the charge density extends over a 200x2 ladder when the sweep number is increased.

Figure 1(b) shows a correlation function 
$\langle S^z(l_x,l_y) n_h(j_0)\rangle$  which measures the spin profile when a dynamic hole 
is on site $j_0$; here $j_0 = (200,2)$ on a $400\times2$ ladder.  
 With  
 this correlation function  shown on a log scale as a function of  distance $l_x$ along the ladder, 
the exponential confinement of the 
 spin and charge is apparent in the   linear $l_x$ dependence.   
 A linear fit 
  gives a decay length of $\xi=3.14$ for $\alpha=1$; this  matches closely with previous results of 3.19(1) for the spin-spin correlation length in the undoped ladder.\cite{rvbladder}  
  (In contrast, Zhu {\it et al.}  
  reported that a similar correlation function decayed as a power law for $\alpha > \alpha_c$.)
\begin{figure}[htbp]
\includegraphics[width=4.2cm]{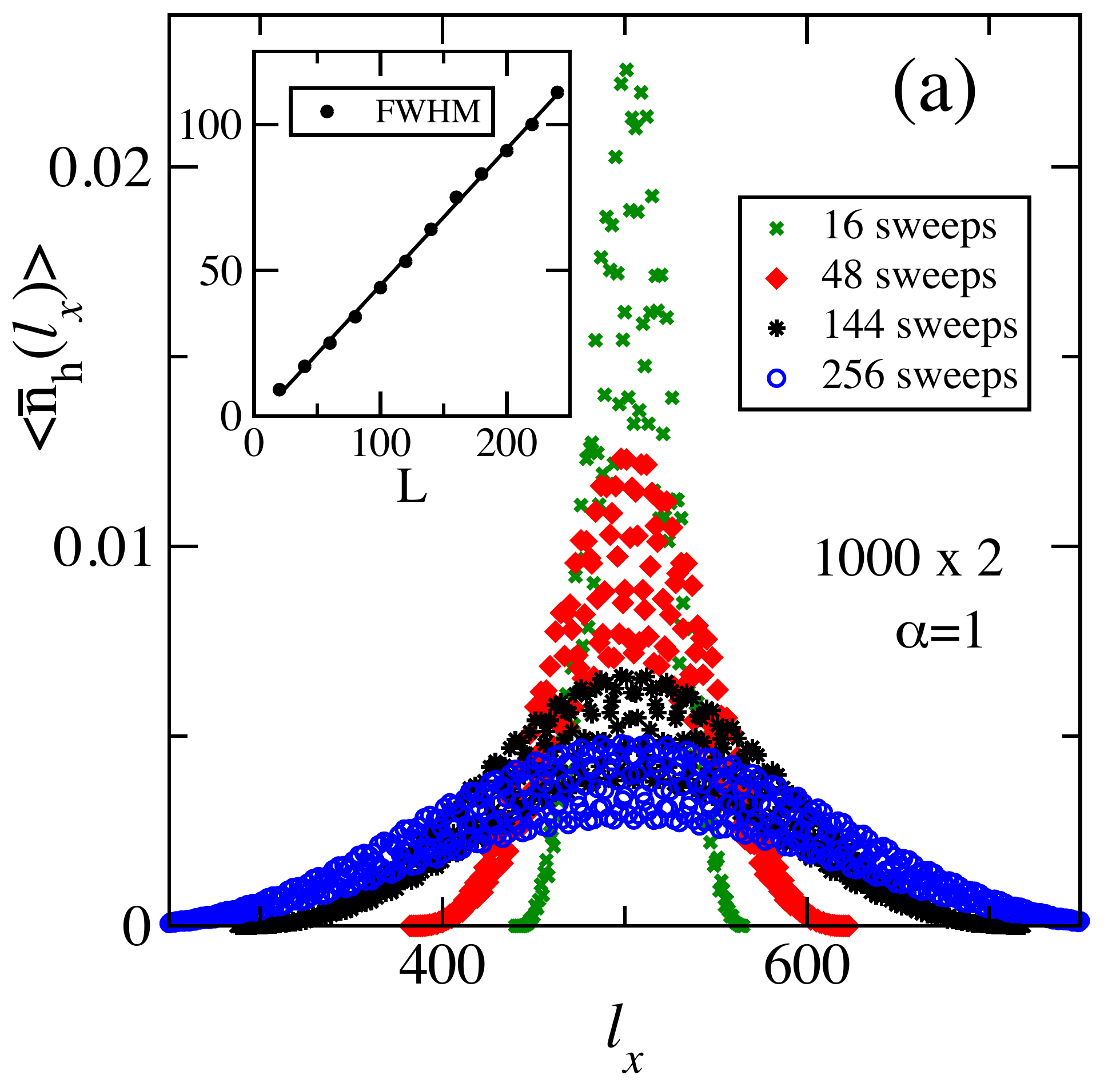}
\includegraphics[width=4.2cm]{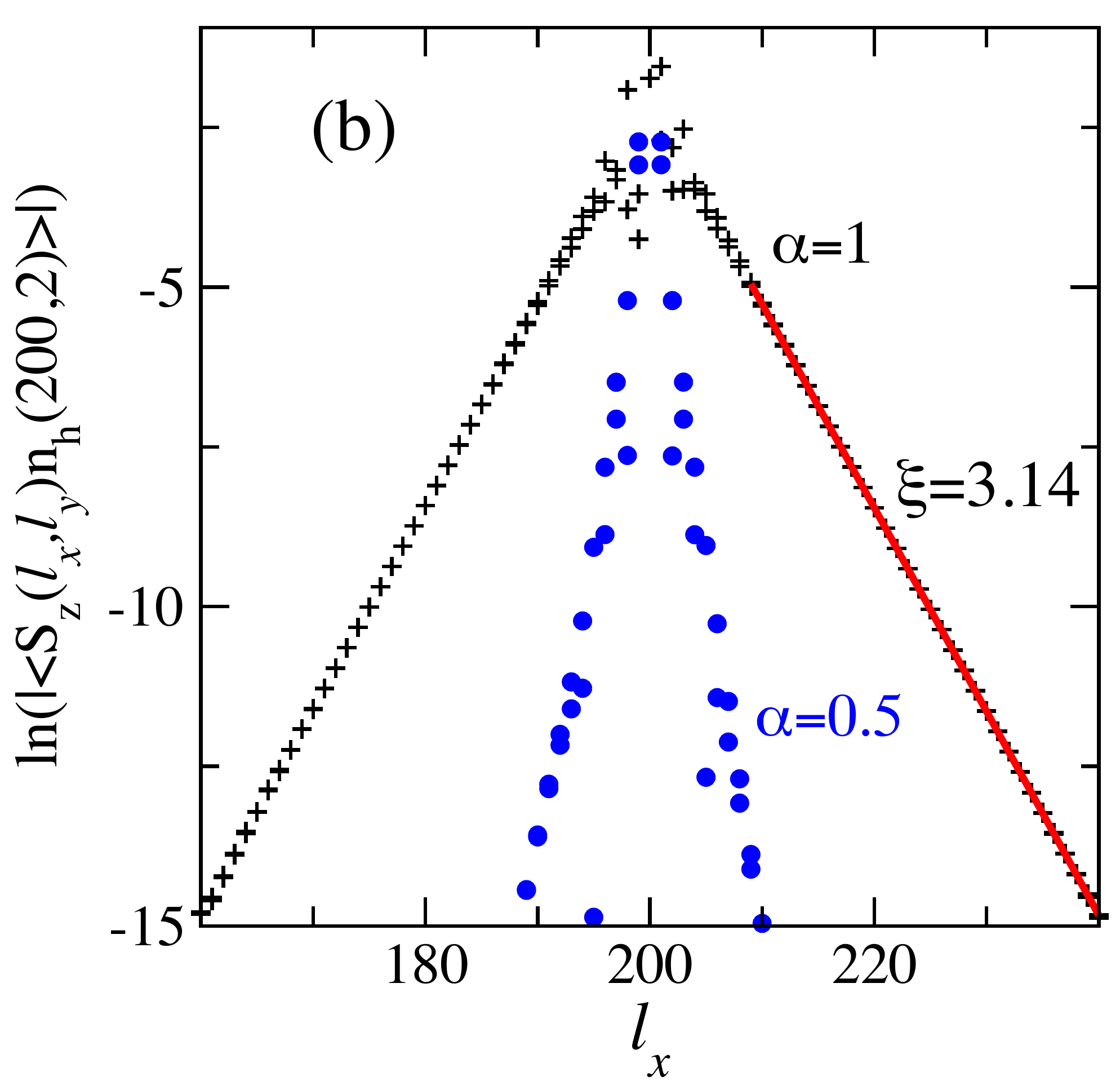}
\caption{(a) The density on each site for $\alpha = 1$ on 
the central portion of a 
$1000\times 2$ system  
for the indicated number of sweeps. 
 The inset shows the full width at half
maximum of the density on a set of smaller lattices 
which were converged in the number of sweeps. (b) A correlation function which measures spin-charge correlations, showing that the spin degrees of freedom are exponentially localized close to a dynamic hole, for $\alpha=0.5$ and $\alpha=1$.  For $\alpha=1$, the red line shows a linear fit to the data.
\label{fig:1}}
\end{figure}

To obtain the one particle spectral function, instead of evolving $|\psi\rangle$
with DMRG sweeps, we evolve it in real time, obtaining the state
\be
|\psi(t)\rangle = \exp(-i t H) c_{j_0,\downarrow} |\phi\rangle
\ee
After each time-step, the Green
 function,  
\be
G(j,t) = \langle \phi | c^\dagger_{j\downarrow} |\psi(t)\rangle e^{i E_0 t} \ ,
\ee
(defined here without the usual $i$ prefactor) was measured for all sites $j$,
where $E_0$ is the ground-state energy of the undoped ladder.  
As time evolves, the wavepacket spreads out.  We always stop the simulation at a time $t_{max}$ before
the packet reaches the edges of the system. Thus any finite size effects are completely
negligible, arising only from the undoped state, which has a 
correlation length that is very small compared to $L$.
Other sources of error are the finite $t_{max}$,
finite truncation error, and finite size of the time steps. Using time steps $\tau=0.05-0.1$, we found 
the time step error was small enough to have no visible effects on any of the figures below.
To measure and control the other two errors, we varied the number of states kept (up to
$m=2000$) and the maximum time (up to $t_{\rm max}=100$).  
Any errors in the results we show primarily appear as slight broadenings of
the spectra, and have no impact on our conclusions.

The ladder is symmetric under reflection symmetry which interchanges the two
legs;  correspondingly, the one-hole states can be classified by their
symmetry, $\Lambda=\pm1$, under reflection.  Similarly, the Bloch wave-number is
a good quantum number.  Thus, to interpret the results physically, we perform
the Fourier transform of $G(j,t)$ with respect to time (using $G(j,-t) = G(j,t)^*$) 
and position along the
ladder, projected onto the space of states of a given reflection symmetry using
both linear prediction\cite{WhiteAffleck} and windowing to deal with a finite
$t_{\rm max}$. The real part of this quantity is the  spectral
function $A(k,\omega)$,
which is shown for $\Lambda=+1$  
in Fig. \ref{fig:2}(a) for the case
$\alpha=1$.  The Supplementary Information
section contains a further discussion of the tDMRG and 
figures of $A(k,\omega)$ for more values of $\alpha$.

The spectral weight is characterized by a sharply defined dispersing pole
separated by a gap of order $J$ from a 
quasi-particle-magnon continuum. For $\alpha=1$,
the minimum in the quasi-particle dispersion occurs at $k_{min}  
\approx 2.01=0.640\pi $. A slice 
of the spectral weight for $\alpha=0.7$ (just above $\alpha_c \approx 0.68$) at $k_{min} 
\approx 2.85=0.907\pi$ is
plotted versus $\omega$ in Fig. 2(b). The dispersion of the pole in the quasi-particle spectrum 
versus 
$k$ for several values of $ \alpha$ is shown in Fig. 2(c).  As $ \alpha $ increases 
beyond $\alpha_c$ , $k_{min}$ moves away from $\pi$ and at large values of $\alpha$ 
approaches $\pi/2$.  
\begin{figure}[htbp]
\includegraphics[width=8.8cm]{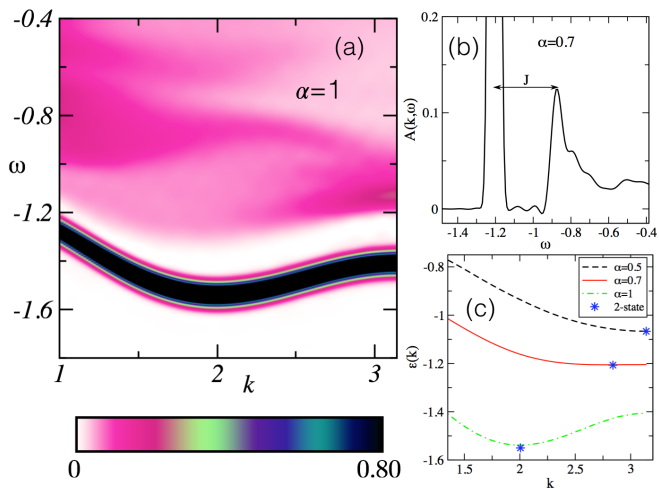}
\caption{(a) Spectral weight function $A(k,\omega)$ in the ground-state
($\Lambda=+1$) reflection parity sector for the $t$-$J$ ladder with $\alpha=1$,
obtained with tDMRG  
color indicating the value of $A(k,\omega)$.  We work in energy units where $t=1$.
(Results for odd reflection parity, $\Lambda=-1$, are shown in the Supplemental Section.)
(b) $A(k,\omega)$ near the
quasiparticle peak for $\alpha = 0.7$ at $k_0/\pi=0.907$.
The gap to the start of the continuum spectrum is of order $J$. (c) Quasiparticle dispersions for
$\alpha=0.5, 0.7, 0.9$, obtained from tDMRG.  The stars show the values of
$k_0$ and $\varepsilon_0$ obtained from separate ground state DMRG calculations.
\label{fig:2}}
\end{figure}

For a given value of $\alpha  $, the minimum hole energy $ \varepsilon_0$  and the corresponding 
wave vector $ k_0 $ can be determined from the dispersion of the peak in $A(k,\omega)$. 
Alternatively, for a given value of $ \alpha $, the energy $\varepsilon_0 $ and wave vector $  k_0 $ 
can be determined directly from our ground state DMRG calculations. The energy minimum
$ \varepsilon_0 $  for a given value of  $ \alpha$  is equal to the difference in
the one hole and zero hole ground state energies. 
The wave vector $ k_0 $ associated with the one-hole ground state can be determined from the 
peak in the spatial Fourier transform of $F(j,s)$, which sharpens as the sweep number $s$ increases.
 Plots of $\varepsilon_0$ and $k_0$ versus $\alpha$ are shown 
in Fig. 3.

\begin{figure}[htbp]
\includegraphics[width=8cm]{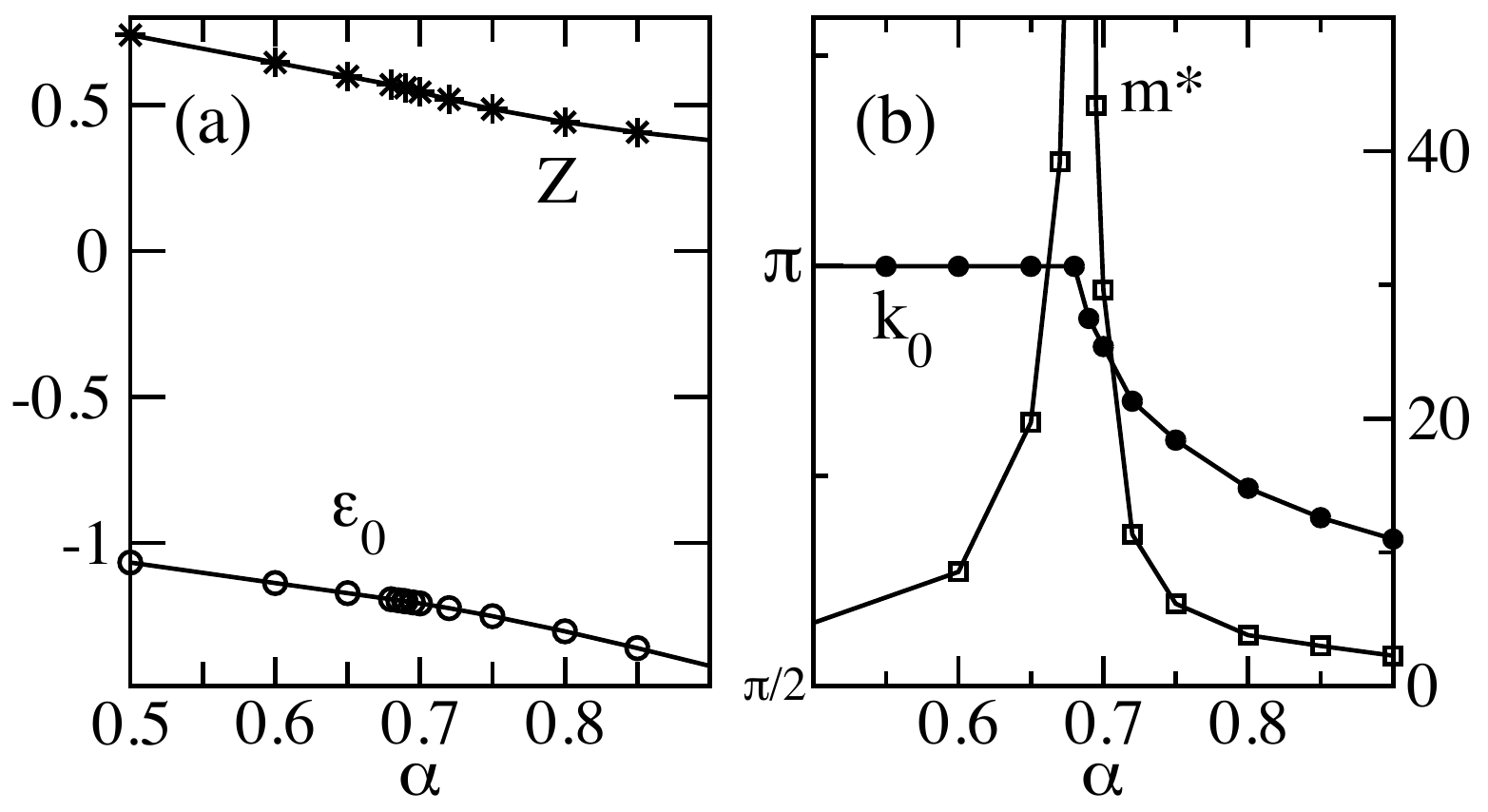}
\caption{The one quasi-hole properties as a function of $\alpha$: The figure shows (a) $Z$ and $\varepsilon_0$,
and (b) $k_0$ and $m*$.  These results were obtained from $400\times 2$ systems from
measurements as the hole spread out with successive sweeps.
\label{fig:3}}
\end{figure}

Similarly, while $ m^* $ can be extracted from the curvature of the quasi-particle dispersion around $ k_0 $ 
  and $Z$ can be obtained from a frequency integration 
 of $A(k_0,\omega)$,  both of these quantities can be directly determined with higher accuracy 
from the ground state DMRG calculations. 
An estimate of the quasi-particle spectral weight $Z$ is given by
\be
Z(s) = \sum_j
| F(j,s)|^2
\ee
We find that this estimate converges very rapidly with the number of sweeps
$s$, much more rapidly than the hole
spreads out.  As the DMRG sweeps continue, 
the energy $E$ of $|\psi(s)\rangle $ converges towards that of the one-hole ground state
with a correction that varies as $ (8m^*\langle x^2\rangle )^{-1} $.  Here, $\langle x^2\rangle  $ is the
variance of the position of the hole, determined from $\langle \bar n_h(l_x) \rangle$.  %
By 
plotting $E$ versus $\langle x^2\rangle ^{-1}$, with each point corresponding to a different sweep, 
one can obtain an estimate of $m^*$.  In addition, one can increase the accuracy of the estimate for $Z$
for the infinite ladder by extrapolating $Z$ versus $\langle x^2\rangle ^{-1}$.
For $\alpha=1$, for example, we obtain $Z = 0.34067(1)$.
Plots of $Z$ and $m^*$ are shown in Fig. 3.  
As seen in this figure, there is a sharp change in the quasi-particle character that
occurs at $\alpha_c = 0.68$. There are kinks in the slopes of $\varepsilon_0,k_0 $ and $Z$ 
and the curvature of the quasi-particle dispersion vanishes giving rise to a
divergence in the effective mass. The shift in $k_0$ away from $\pi$ gives rise to the
oscillations in the charge density, as has been previously noted by Zhu {\it et al}, 
which are found to occur at wave-number $2k_0$.

Since the one hole state has a well defined quasi-particle spectral weight, 
many properties that are measurable in numerical experiments on systems 
with large but finite L can be understood in terms of the simpler problem of
 one-hole on a 2-leg band insulator.
   Central to this understanding is the quasi-particle dispersion relation
which determines  the values of $k=\pm
k_0$  at which $\varepsilon(k)$ is minimized, and the dependence of the hole-energy near this point,
$\varepsilon(k) =E_0+\varepsilon_0 +
(k-k_0)^2/2m^\star +\ldots$, where $m^\star$ is the effective mass. 
In order to minimize its zero-point energy 
on a ladder of large but finite
length $L$, the one-quasiparticle ground-state will always spread to fill the
extent of the ladder,
\be
\psi_L(n,\tau) \sim \sin(\pi n/L) \cos(k_0n-\theta),
\ee
where $\theta=k_0L/2$.  The minimum in the  ground state energy of the one hole state is 
\be
\varepsilon(L) =E_0+ \varepsilon_0 + \pi^2/(2m^\star L^2) + \ldots .
\label{box}
\ee 
Since integrating out the gapped spin degrees of freedom 
inevitably renormalizes the bare dispersion, for comparison purposes we consider a non-interacting model with  
band structure
\be
E(k) = - \Lambda t_\perp -2t_\parallel\cos(k) - 2t_\parallel^\prime \cos(2k)
\ee
 in which $\Lambda=\pm 1$ correspond to the valence and conduction bands, respectively, all the $t$'s are assumed non-negative and
   the rung hopping parameter $t_\perp$ 
   to be sufficiently large compared to the near-neighbor and
next-near-neighbor leg hopping parameters $t_\parallel $ and
$t_\parallel^\prime $ that the undoped system has an insulating gap.  This
dispersion is similar to that shown in Fig. 2c. The parameter that plays a role
analogous to $\alpha$ is $\tilde \alpha \equiv 4 t_\parallel^\prime/t_\parallel$;  for $0\leq \tilde \alpha \leq 1$,
 the top of
the valence band occurs at $k=\pi$, while for $\tilde \alpha > 1$, the top of
the valence band occurs at $k=\pm k_0$ where $\cos(k_0) = -1/\tilde \alpha$.
The critical dependences of $ \varepsilon_0=-E(k_0)$ , $k_0$ and $m^{*} $ on $ \tilde\alpha $
can be readily derived from the band dispersion Eq. (7). 

\be
\frac {[\varepsilon_0 + t_\perp]} {t_\parallel} =
\left\{
\begin{array}{ccc}
 -{(4-\tilde \alpha)}/{2} &  {\rm for}  & \tilde \alpha< 1  \\
 -{(2+\tilde \alpha^2)}/{2\tilde \alpha}
& {\rm for}   &   \tilde \alpha > 1
\end{array}
\right. 
\ee

\be
\frac {1}{t_\parallel} \frac {d \varepsilon_0}{d\tilde\alpha}=
\left\{
\begin{array}{ccc}
1/2 &  {\rm for}  & \tilde \alpha< 1  \\
(2-\tilde \alpha^2)/2\tilde\alpha^2
& {\rm for}   &   \tilde \alpha > 1
\end{array}
\right. 
\ee

  \be
\label{qofx}
\pi - k_0= 
\left\{
\begin{array}{ccc}
 0 &  {\rm for}  & \tilde \alpha< 1  \\
 \sqrt{
 {2(\tilde \alpha-1)}/ {\tilde \alpha}} 
  & {\rm for}  &  1 \gg (\tilde \alpha-1) >0 \\
\pi/ 2 - 
{1}/ {\tilde \alpha} 
& {\rm for}   &   \tilde \alpha \gg 1
\end{array}
\right. 
\ee
and
\be
\label{mstar}
 {m^*} =\frac 1 {2t_{\parallel}}
\left\{
\begin{array}{ccc}
  [1-\tilde \alpha]^{-1} &  {\rm for}  & \tilde \alpha <1  \\
{ \tilde\alpha
}{(\tilde \alpha^2 - 1)^{-1}} &  {\rm for}   &   \tilde\alpha > 1
\end{array}
\right. 
\ee 
The qualitative features 
observed in the evolution of the one-hole state  
  of the $t-J-\alpha$ model as a function of  $\alpha$
are 
reflected in the band model as a function of $\tilde\alpha $. i) The one-hole energy   $  \varepsilon_0 $  has
a non-analytic change in  slope  at $\tilde\alpha=\tilde\alpha_c $  given
by Eq. [9]. ii) The vector $ k_0(\tilde\alpha)$ has a square-root singularity
at $\tilde\alpha= \tilde\alpha_c$ as given by Eq. [10], and $2k_0$  determines the oscillations of the charge density. 
 iii) The effective  mass $m^{*}(\tilde\alpha)$ diverges linearly upon approaching $\tilde\alpha_c$
from both sides as given in Eq. [11].

In the Supplemental Information we make this connection formal:  We define a
$t-J$-Hubbard model Hamiltonian that in one limit is equivalent to the
$t-J-\alpha$ model of Eq. 1, and in another limit represents a non-interacting
band-insulator, with the band structure given in Eq. (7).  Our DMRG results
establish that there is no gap-closing 
 and so  no barrier to adiabatic
continuity 
 upon reducing the model to one of decoupled rungs in the $\alpha=0$
limit.  In this limit the interactions can be adiabatically set to 0, again
without any gap closures.  Finally, in the solvable non-interacting limit, we
restore the hopping matrix elements along the ladder, $t_\parallel$ and
$t_\parallel^\prime$, still without encountering any gap closures.  (The final
two steps are readily studied analytically.)  This analysis constitutes a proof
that the low energy one-hole states of the $t-J-\alpha$ model are adiabatically
connected to those of a non-interacting band insulator which holds regardless
of the value of $\alpha$ in the entire range we have studied.

We would like to thank A.L.~Chernyshev, 
Zheng Zhu, and Hong-Chen Jiang
for insightful discussions. SRW acknowledges support from
the NSF under grant DMR-1161348 and from the Simons Foundation through the Many Electron Collaboration. 
DJS acknowledges the support of the Center for 
Nanophase Materials
ORNL, which is sponsored by the Division of Scientific User Facilities, U.S. DOE.
SAK acknowledges support from the NSF under grant DMR-1265593.

\appendix

\section {Supplemental material}

The fact that the low energy one-hole states of  the two-leg $t-J$ ladder can
be adiabatically connected to the corresponding states of  a two-leg
non-interacting semiconductor is established by explicit construction.  In
addition, details of the time dependent DMRG calculations are given, and
additional plots of the spectral function are presented.  


\subsection{Adiabatic continuity to the noninteracting limit}

{\it The $t-J$ Hubbard model:}  Consider the Hamiltonian for electrons in a two leg ladder with 
 sites labeled by the
  leg index $\tau=u,\ d$ and rung index $j$:
\bea
&&H(t_\parallel,t_\parallel^\prime,t_\perp,J,J_\perp,U)\equiv \nonumber  \\
&&\ \ -\sum_{j,\tau \sigma}\left[t_\parallel c_{j,\tau,\sigma}^\dagger c_{j+1,\tau,\sigma} + t_\parallel^\prime c_{j,\tau,\sigma}^\dagger c_{j+2,\tau,\sigma} + {\rm H.C.}\right]\nonumber  \\
&&\ \ -\sum_{j,\sigma}\left[t_\perp c_{j,u,\sigma}^\dagger c_{j,d,\sigma} + {\rm H.C.}\right] \nonumber \\
&&\ \ + J_\parallel\sum_{j,\tau}\left[ \vec S_{j,\tau}\cdot \vec S_{j+1,\tau} -\frac 1 4 n_{j,\tau}n_{j+1,\tau}\right ]\nonumber \\
&&+J_\perp\sum_{j,\tau}\left[  \vec S_{j,u}\cdot\vec S_{j,d}-\frac 1 4 n_{j,u}n_{j,d} \right]  \nonumber \\
&&\ \ +U\sum_{j,\tau}\left[ c_{j,\tau,\uparrow}^\dagger  c_{j,\tau,\downarrow}^\dagger c_{j,\tau,\downarrow}c_{j,\tau,\uparrow}\right]
\label{H}
\eea 
where $t_\parallel$ and $t_\parallel^\prime$ are the first and second neighbor hopping along the ladder, $t_\perp$ is the hopping between rungs, $J_\parallel$ and $J_\perp$ are the corresponding exchange couplings, $c_{j,\tau,\sigma}^\dagger$ creates an electron with spin-polarization $\sigma$ on site $(j,\tau)$,  
\be
\vec S_{j,\tau} = \sum_{\sigma,\sigma^\prime} c_{j,\tau,\sigma}^\dagger \vec \tau_{\sigma,\sigma^\prime} c_{j,\tau,\sigma^\prime} 
\ee 
is the spin  and
\be
\vec n_{j,\tau} = \sum_{\sigma} c_{j,\tau,\sigma}^\dagger c_{j,\tau,\sigma} 
\ee 
is the electron density on site $(j,\tau)$.  In contrast to the $t-J$ model, this hamiltonian   acts on  the full fermionic Hilbert space in which there is no constraint on double-occupancy  sites, although this constraint can be obtained dynamically by taking the limit in which the  on-site Hubbard repulsion $U$ tends to $\infty$.
Thus, the $t-J$ model is simply the $U\to\infty$ limit of this model;  specifically, the version of the $t-J$ model studied by Zhu {\it et al} (Eq. (1) of our paper) is
\be
H_{t-J} \equiv \lim_{U\to\infty} H(\alpha t,0,t,\alpha J,J,U) .
\label{tJ}
\ee
with $t=t_\parallel$, $J=J_\parallel$, and $\alpha=t_\parallel/t_\perp=J_\parallel/J_\perp$.
On the other hand, in contrast to the $t-J$ model, this model has a non-interacting limit,
\be
\label{non}
H_{non}(t,t^\prime,t_\perp) = H(t,t^\prime,t_\perp,0,0,0).
\ee

{\it Non-interacting limit:}
The band-structure of the non-interacting two-leg ladder described  by $H_{non}$, is trivially obtained using Bloch's theorem and taking advantage of the reflection symmetry which exchanges the two legs.  The dispersion relation for this problem is
\be
\varepsilon_\gamma(k) = -\gamma t_\perp -2t_\parallel\cos(k) - 4t_\parallel^\prime \cos^2(k) +2t_\parallel^\prime
\ee
 where $\gamma=\pm 1$ is the reflection symmetry and $k$ is the Bloch wave-number. We impose the condition the system be insulating when there is one electron per site by considering  $|t_\perp|$ is sufficiently large compared to $|t_\parallel |$ and $|t_\parallel^\prime |$, so that there is a gap in this spectrum.  To be explicit, we further restrict consideration to the case in which all the $t$'s are non-negative. The parameter that plays a role analogous to $\alpha$ is $\tilde \alpha \equiv t_\parallel^\prime/4t_\parallel$;  for $0\leq \tilde \alpha \leq 1$, the top of the valence band occurs at $k=\pi$, while for $\tilde \alpha > 1$, the top of the valence band occurs at $k=\pm k_0$ where $\cos(k_0) = 1/\tilde \alpha$.

{\it Adiabatic continuity:}  In a quantum system with a gap, the notion of adiabatic continuity can be given a precise definition -- the states of two systems are adiabatically connected if it is possible to continuously deform the Hamiltonian in such a way that the gap never closes in turning it from that of the initial to the final system.
  If we restrict ourselves (as we often do) to adiabatic paths that preserve
certain symmetries, then two states with different symmetry related quantum
numbers can never be adiabatically connected.  Conversely, we can study
adiabatic continuity within a given subspace of Hilbert space specified by
these quantum numbers, even if somewhere along the path there might occur a
region where the absolute ground-state lies in a different subspace.\cite{thermo}

{\it An adiabatic route from $H_{t-J}$ to $H_{non}$:}  Here we vary the parameters in $H(t_\parallel,t_\parallel^\prime,t_\perp;J_\parallel,J_\perp;U)$ to trace an adiabatic path from $H_{t-J}$ to $H_{non}$ always preserving translational (with  periodic boundary conditions), reflection, spin rotational, and gauge  (number conservation) symmetries:
\bea
\label{adiabatic}
&&H_{t-J} = H(\alpha t,0,t;J,\alpha J_\perp;U=\infty)  \\
&&\longrightarrow  H(0,0,t_\perp;0,J_\perp;U=\infty) \Longrightarrow H(0,0,t_\perp;0,0;0) \nonumber\\
&& \Longrightarrow H(t_\parallel,-t_\parallel^\prime,t_\perp;0,0;U=0) = H_{non}(t_\parallel,t_\parallel^\prime,t_\perp).
\nonumber
\eea
The single-arrow represents steps  for which the existence of a non-zero gap along the entire path and hence the possibility of adiabatic evolution has been established using DMRG results for the $t-J$ ladder, while the double-arrows represent steps that can be justified analytically.

i) In the first step, the system is deformed into a  set of decoupled rungs.  
That there is a (spin) gap in the excitation spectrum of the undoped ladder
over the entire pertinent range of $\alpha$  is well established by the present
DMRG calculations as well as by those of Zhu {\it et al}\cite{Zhu1,Zhu2,Zhu3,Zhu4}.
For one doped hole, we first restrict attention to the
subspace with $k=k_0$ and $\lambda=1$ -- {\it i.e.} the sector which contains
the one-hole ground-state for the particular initial value of $\alpha=\alpha_0$
being considered.  (For $\alpha_0< \alpha_c$ this is $k_0=\pi$, whereas for
$\alpha_0>\alpha_c$ this is an appropriate smaller value of $k_0$.)  What is
apparent from our DMRG results is that for the entire range $\alpha$ we have
studied, there is a gap of order $J$ separating the quasi-hole state from the
multi particle continuum.  Notice that although in the case that $\alpha_0<
\alpha_c$, the one-hole ground-state remains at $k=\pi$ for the entire range of
$\alpha$ between $\alpha=\alpha_0$ and $\alpha=0$, for $\alpha_0 > \alpha_c$,
the ground-state sector changes as $\alpha$ varies from $\alpha=\alpha_0$ to
$\alpha=\alpha_c$.  This does not, however, act as a barrier to adiabatic
evolution, since as long as we maintain translational symmetry, we are free to
restrict our attention to the subspace with $k=k_0(\alpha_0)$ for the entire
process.

ii)  Once the system consists of decoupled dimers, the spectrum can be readily computed 
analytically and it is easy to see that it is possible to simultaneously
decrease the values of $U$ and $J_\perp$, without ever closing the gap, to the
point at which the system consists of non-interacting electrons confined to the
bonding states on each rung.  Note that throughout this portion of the
evolution, the one-hole ground-state is $2L$ fold degenerate, but in any sector
specified by Bloch wave-vector, $k$, and spin polarization, $\sigma$, the
one-hole ground-state remains non-degenerate.

iii)  Once all interactions have been quenched, it is simple to compute the band-structure for arbitrary $t_\perp$, $t_\parallel$, and $t_\parallel^\prime$.  For sufficiently large  $t_\perp$, the semiconducting gap of the undoped system the gap at fixed $k$ in the presence of one doped hole remain non-zero throughout this process.    If in the starting Hamiltonian, $\alpha < \alpha_c$, then we can insure that the one-hole ground-state occurs in the appropriate $k_0=\pi$ sector by ending with a value of $|t_\parallel^\prime| < |t_\parallel|/4$.  If the starting Hamiltonian has $\alpha > \alpha_c$, then by ending the adiabatic evolution with $t_\parallel^\prime=-t_\parallel/4\cos(k_0)$, we reach a situation in which the lowest energy one-hole state occurs at $k=k_0(\alpha)$, {\it i.e.} in the same sector of Hilbert space as in the initial Hamiltonian.

{\em This constitutes the proof that the two-leg $t-J$ ladder is adiabatically
connected to a band-insulator, both for the undoped system and in the presence
of one doped hole.}

\

\subsection{Time dependent DMRG and spectral functions}

The tDMRG results were obtained using a Trotter decomposition\cite{tdmrg1,tdmrg2,tdmrg3}, applying only
nearest neighbor gates, using a reordering of the sweep path through the lattice 
to make this possible.
The initial path was chosen as $(x,y) = (1,1), (1,2), (2,2), (2,1), (3,1)$, \ldots.
Along this path, only bonds on leg 1 are not nearest neigbhor.
We alternate this path with
$(1,2), (1,1), (2,1), (2,2), (3,2)$, \ldots, where leg 2 bonds 
are not nearest neighbor. Write the Hamiltonian as 
$H = H_1 + H_2$
where 
\be
H_1 = \frac{1}{2}H_{\rm rungs} +  H_{\rm leg-1}
\ee
and 
\be
H_2 = \frac{1}{2}H_{\rm rungs} +  H_{\rm leg-2}
\ee
We perform a half sweep, applying bond time evolution operators $\exp(-i \tau
H_{\rm bond})$, applied only on the terms in $H_1$, where $\tau$ is the time step. 
Then, we perform a
half sweep which switches the path to the second ordering, consisting of
applying swap operators on each rung\cite{Stoudenmire}. After this, we perform a half sweep using
the terms of $H_2$.   Then this entire
sequence of three half sweeps is applied entirely in reverse, returning the
sites to their original order.  The reversal also cancels the lowest order
Trotter error, resulting in an overall Trotter error of $O(\tau^2)$ (per unit time), in a time-step 
that progresses by $2 \tau$.

\begin{figure}[htbp]
\includegraphics[width=8cm]{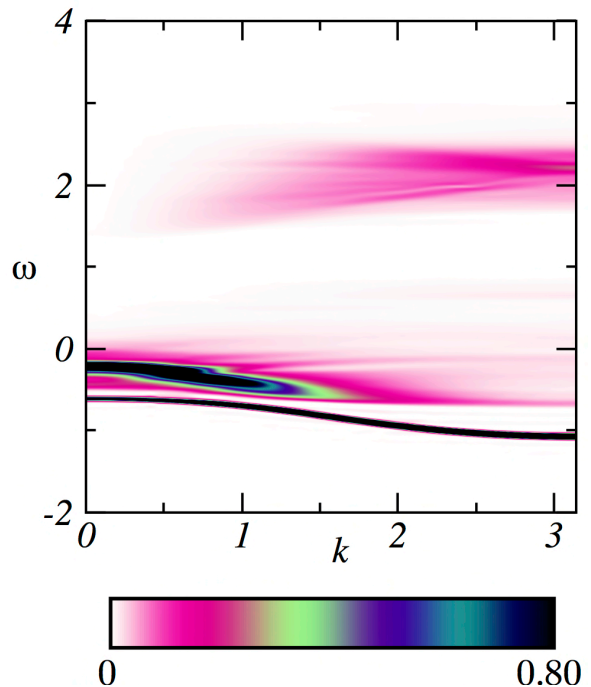}
\caption{ Spectral function for $\alpha = 0.5$ for $\Lambda = +1$. 
The maximum number of states was $m=500$.
\label{fig:1}}
\end{figure}

A Fourier time-space Fourier transform
of $G(j,t)$ yields $A(k,\omega)$, using both linear prediction\cite{WhiteAffleck} and 
windowing to deal with a finite $t_{\rm max}$.
As time evolves, the wavepacket spreads out.  We always stop the simulation before
the packet reaches the edges of the system. Thus any finite size effects are completely
negligible, arising only from the undoped state, which has a very short correlation length.
Other sources of error remain, namely finite maximum time, 
finite truncation error, and finite time step. Using $\tau=0.05-0.1$, we found 
the time step error was small enough to have no visible effects on any of the figures below.

\begin{figure}[htbp]
\includegraphics[width=8cm]{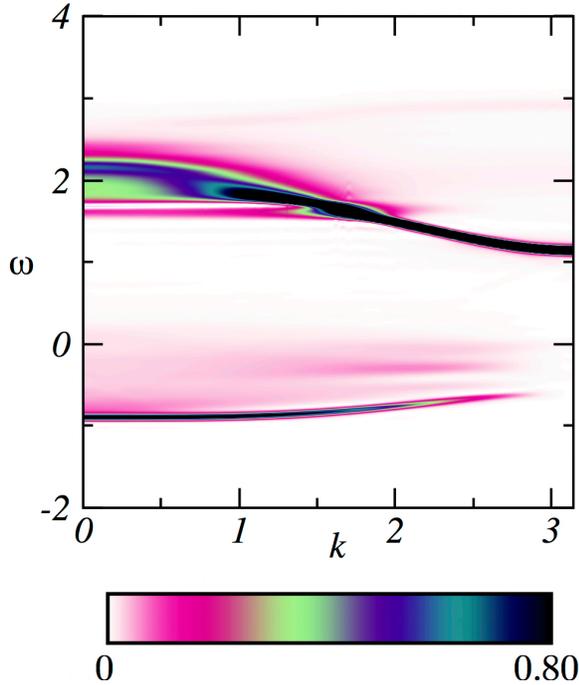}
\caption{ Spectral function for $\alpha = 0.5$ for $\Lambda = -1$, for the same run as in the previous figure.
\label{fig:2}}
\end{figure}

As $t$ increases, the entanglement of the state increases.  If we keep a variable number of states,
specifying a particular truncation error at each step, then the number of states will increase as
time increases. For example, with $\alpha=1$ if we specify a truncation error of $10^{-7}$, the number
of states kept rises as a rapidly increasing function that reaches $m=3000$ at about $t \sim 14 - 15$.
(The entanglement growth is smaller for smaller $\alpha$.)
There are several ways to deal with this entanglement increase: we discuss three approaches. 

\begin{figure}[htbp]
\includegraphics[width=8cm]{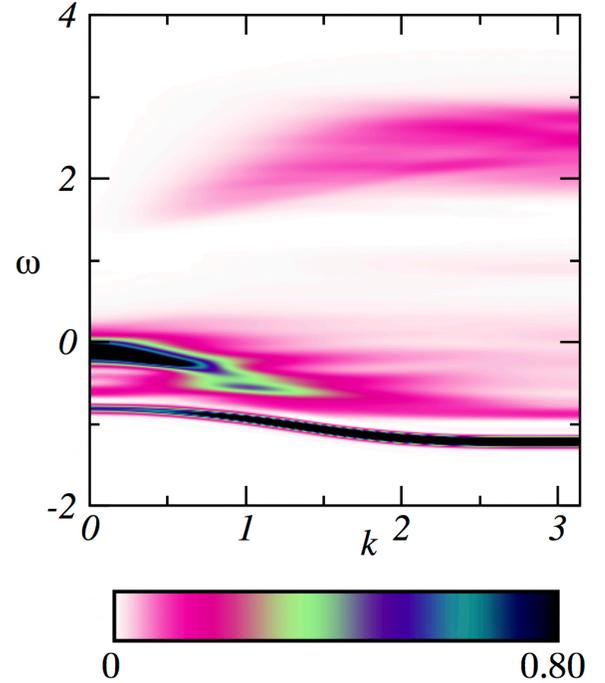}
\caption{ Spectral function for $\alpha = 0.7$ for $\Lambda = +1$.
The maximum number of states was $m=500$.
\label{fig:3}}
\end{figure}

1) One can stop the simulation when
$m$ reaches a cutoff, e.g. stopping at $t_{\rm max} \sim 15$ when $m=3000$ for $\alpha=1$.
One can rely on the linear prediction to extend $t_{\rm max}$ before Fourier transforming.  We did not follow this
approach.  

\begin{figure}[htbp]
\includegraphics[width=8cm]{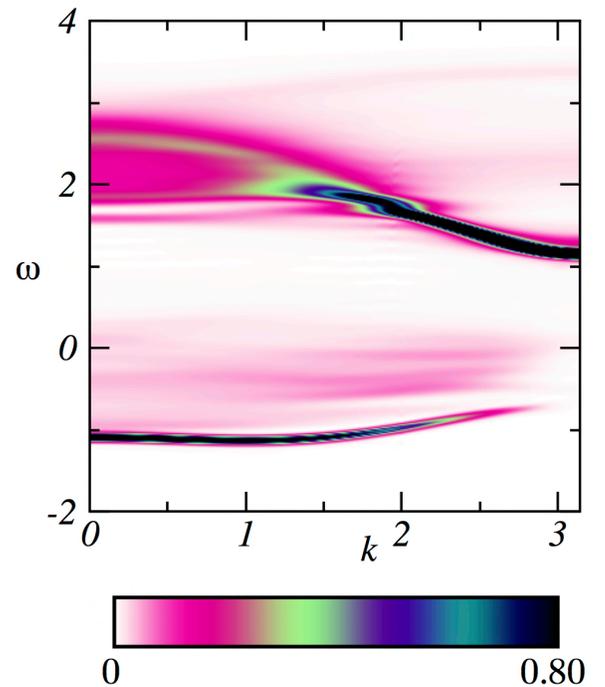}
\caption{ Spectral function for $\alpha = 0.7$ for $\Lambda = -1$, for the same run as in the previous figure.
\label{fig:4}}
\end{figure}

2) One can specify a maximum $m$ to overrule the specified truncation error, giving 
a larger truncation error for larger times.  An advantage of this method is that the increased truncation 
somewhat resembles a windowing function, in that it reduces $G(t)$ as $t$ increases. A windowing function must
be applied anyway, so this is not a very serious error. 
The decrease in $G(t)$ is
not uniform across frequencies--the higher energy states are more entangled, and their amplitude decreases
more rapidly than the low energy states.  This allows good resolution of the quasiparticle part of the spectrum,
and we have generally followed this.  
(One must not ``fix'' the
normalization of the wavefunction after the truncation error--this increases the amplitudes of the low energy
part of the spectrum to fix the loss at high energies, producing poor results.) 
One can vary $m$, $t_{\rm max}$, etc. and check for convergence of the results.  We have mostly followed this procedure.

\begin{figure}[htbp]
\includegraphics[width=8cm]{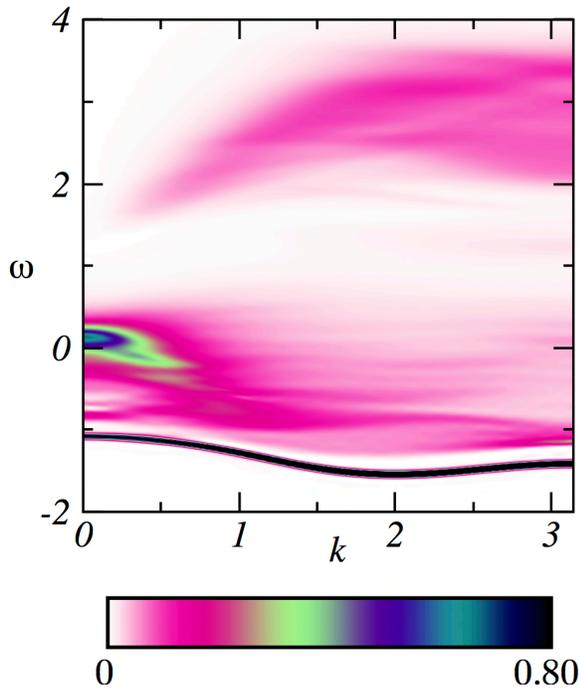}
\caption{ Spectral function for $\alpha = 1.0$ for $\Lambda = +1$, keeping $m=2000$ states.
\label{fig:5}}
\end{figure}

3) One can evolve in imaginary time a fixed distance $\beta$ (not an actual temperature), say $\beta \sim 1$,
before starting the real time evolution.  This diminishes the high energy parts of the state, and
the low energy part that is left has lower entanglement growth.  After the simulation, and after the
standard linear prediction, windowing, and Fourier transforming, one corrects for the initial imaginary time
evolution by replacing $A(k,\omega)$ by $A(k,\omega) e^{\beta \omega}$.  This method works quite well.
For large $\omega$, the results can be poor, because errors are amplified, but it gives a well controlled way to 
zoom in on the low energy part of the spectrum with high accuracy.  This approach can be combined with method 2), using a maximum
$m$.  This method was used  to obtain the spectrum shown in Fig. 2(b).

We present here the spectral functions for several values of $\alpha$, all obtained with method 2). 
The low energy quasiparticle bands are very reliable, and their finite frequency width is a consequence of
finite $t_{\rm max}$.  The higher energy parts have broad features and
also smaller amplitude features, e.g. subtle color variations.  The broad features and distribution of spectral weight are very reliable,
but it can be hard to tell if some of the small amplitude high frequency features are artifacts due
to noise or ringing, without further study, comparing spectra with different accuracy parameters, 
which we have not done very thoroughly.

\begin{figure}[htbp]
\includegraphics[width=8cm]{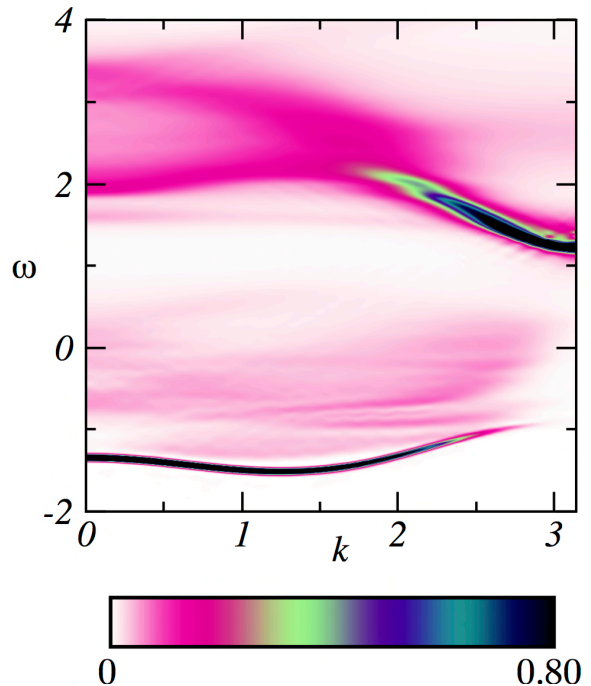}
\caption{ Spectral function for $\alpha = 1.0$ for $\Lambda = -1$, for the same run as in the previous figure.
\label{fig:6}}
\end{figure}



\begin{thebibliography}{99}
\bibitem{thermo}  The thermodynamic definition of adiabaticity is somewhat more
complicated, involving the possibility of a quasi-static evolution at constant
entropy.  Since the one-hole state is not a sensible notion in the
thermodynamic limit, and as in any case the quantum mechanical definition of
adiabaticity is more restrictive than the thermodynamic one, we  deal here
solely with the former.
\end{thebibliography}
\end{document}